# Resource-bounded Dimension in Computational Learning Theory

Ricard Gavaldá [*], María López-Valdés[†], Elvira Mayordomo [†], N. V. Vinodchandran [‡]

**Abstract**

This paper focuses on the relation between computational learning theory and resource-bounded dimension. We intend to establish close connections between the learnability/nonlearnability of a concept class and its corresponding size in terms of effective dimension, which will allow the use of powerful dimension techniques in computational learning and viceversa, the import of learning results into complexity via dimension. Firstly, we obtain a tight result on the dimension of online mistake-bound learnable classes. Secondly, in relation with PAC learning, we show that the polynomial-space dimension of PAC learnable classes of concepts is zero. This provides a hypothesis on effective dimension that implies the inherent unpredictability of concept classes (the classes that verify this property are classes not efficiently PAC learnable using any hypothesis). Thirdly, in relation to space dimension of classes that are learnable by membership query algorithms, the main result proves that polynomial-space dimension of concept classes learnable by a membership-query algorithm is zero.

## 1 Introduction

Computational learning theory studies the performance obtained and the resources needed in machine learning. This formalization dates back to the work of Valiant that in 1984 introduced the Probably Approximately Correct (PAC) learning model [27]. This model has been explored and several alternatives have been considered such as the query learning model by Angluin [3] or the on-line mistake-bound learning model by Littlestone [19]. The main open problems in computational learning theory concern the limits of each learning model. We want to prove that a certain class of concepts is not learnable under a certain model, therefore establishing a lower bound in the inherent learning complexity of that class. On the other hand the quest for new and efficient learning algorithms is a very active and challenging area. This paper explores the relationship between computational learning theory and effective dimension with the ultimate goal of obtaining nonlearnability results from dimension results as well as translating learning algorithms into effective dimension proofs.

[*]Departament de Llenguatges i Sistemes Informàtics, Universitat Politècnica de Catalunya. 08034 Barcelona, SPAIN. gavalda@lsi.upc.edu.

[†]Departamento de Informática e Ing. de Sistemas, Instituto de Investigación en Ingeniería de Aragón, Universidad de Zaragoza. 50018 Zaragoza, SPAIN. marlopez@unizar.es, elvira@unizar.es This research was supported by Spanish Government MEC projects TIN2005-08832-C03-02, TIN2008-06582-C03-02.

[‡]Dept. of Computer Science & Engineering, University of Nebraska-Lincoln. Lincoln, NE 68588, USA. vinod@cse.unl.edu

Resource-bounded dimension (or effective dimension) was developed by Lutz [22], as a way to overcome limitations of resource-bounded measure [21, 20]. Both of them are quantitative tools in Computational Complexity that were introduced in order to analyze the size of complexity classes. Their power is witnessed by an interesting list of results both in Computational Complexity and Information Theory (see [11, 10] for an updated bibliography). The main antecedent of this paper is the work of Watanabe et al that investigated the resource-bounded measure of classes that are learnable by PAC or equivalence query algorithms [18]. More specifically, they proved that $i$) P/poly subclasses that can be learned with PAC algorithms have polynomial measure 0 if EXP $\nsubseteq$ AM; and $ii$) the P/poly subclasses that can be learned with membership queries have polynomial measure 0. From these results, hypotheses were provided on the resource-bounded measure of circuits that imply the non-learnability of Boolean Circuits in polynomial time.

On the other hand, in the context of effective dimension, Hitchcock explored the relationship of dimension with logarithmic loss unpredictability in [12].

This paper provides new results in line with those cited above. Firstly, in relation to on-line mistake-bound learning, we obtain an upper bound of the polynomial-time dimension of concept classes that are learnable by on-line algorithms in exponential time and with $\alpha 2^n$ mistakes. Moreover, we prove that this upper bound is optimal. Based on our results Hitchcock [14] has further investigated the case of subexponentially many mistakes and null dimension, with interesting applications [7]. Secondly, in relation with PAC learning, we show that the polynomial-space dimension of PAC-learnable classes of concepts is zero. This provides a hypothesis on effective dimension that implies the inherent unpredictability of concept classes, an interesting property in computational learning that corresponds to classes not efficiently PAC learnable using any hypothesis. Furthermore, there exist connections between hardness results for PAC learning and constructions in the field of public-key cryptography [16] that can now be rephrased with effective dimension hypothesis. Finally, it is studied the space dimension of classes that are learnable by membership query algorithms. The main result proves that polynomial-space dimension of concept classes learnable by a membership-query algorithm is zero. This can be used to demonstrate that, for classes that are complex in the dimension sense, a complex representation is necessary in order to efficiently learn the class with membership queries.

## 2 Preliminaries

A *string* is a finite and binary sequence $w \in \{0,1\}^*$. The *Cantor space* **C** is the set of all infinite binary sequences. Let $|w|$ denote the length of the string $w$ and $\lambda$ denote the empty string. Let $x[i \ldots j]$ for $0 \leq i \leq j$ denote the $i$-th through the $j$-th bits of $x$, where $x \in \{0,1\}^* \cup \mathbf{C}$. Let $wx$ denote the concatenation of the string $w$ and the string or sequence $x$. Let $w \sqsubseteq x$ denote that $w$ is a *prefix* of $x$. Let $s_0, s_1, s_2 \ldots$ be the enumeration of $\{0,1\}^*$ in lexicographical order and $s_0^n, s_1^n \ldots s_{2^n-1}^n$ be the enumeration of $\{0,1\}^n$. Each language $L \subseteq \{0,1\}^*$ can be identified with its characteristic sequence $\chi_L \in \mathbf{C}$ where

$$\chi_L[i] = \begin{cases} 1 & \text{if } s_i \in L. \\ 0 & \text{if } s_i \notin L. \end{cases}$$

Abusing notation, $L$ can be seen as a language or as a characteristic sequence. Also, $L^{=n}$ can be seen as $L \cap \{0,1\}^n$ or as the sequence $L[2^n - 1 \ldots 2^{n+1} - 2]$. To solve the resulting ambiguity it will be used $\#A$ for the cardinality of a set $A$.

Let $\Delta$ denote any of the following classes of total functions,

$$\begin{aligned}
\text{pspace} &= \{f : \{0,1\}^* \to \{0,1\}^* \mid f \text{ is computable in polynomial space}\} \\
\text{p} &= \{f : \{0,1\}^* \to \{0,1\}^* \mid f \text{ is computable in polynomial time}\} \\
\text{p}_2 &= \{f : \{0,1\}^* \to \{0,1\}^* \mid f \text{ is computable in } n^{(\log n)^{O(1)}} \text{time}\} \\
\text{plogon} &= \{f : \{0,1\}^* \to \{0,1\}^* \mid f \text{ is computable by an on-line machine with} \\
&\quad \text{working and output space polylogarithmic in the size of the input}\}
\end{aligned}$$

Let $D$ be a discrete domain such as $\mathbb{N}$ or $\{0,1\}^*$, a function $f : D \to [0,\infty)$ is $\Delta$-*computable* if there exists a function $\widehat{f} : D \times \mathbb{N} \to \mathbb{Q} \cap [0,\infty)$ in $\Delta$ such that for all $(w,n) \in D \times \mathbb{N}$, $|\widehat{f}(w,n) - f(w)| \leq 2^{-n}$ (with $n$ coded in unary and the output coded in binary).

## 2.1 Effective dimension

Effective dimension was introduced by Lutz as a generalization of the classical Hausdorff dimension and is defined in terms of $s$-gales [22, 13]. Effective dimension initial purpose was to serve as a quantitative tool in Computational Complexity, distinguishing complexity classes by size, but lately several interesting applications have been explored, including connections to Information Theory and compression algorithms [15, 24].

**Definition.** Let $s \in [0, \infty)$, an $s$-*gale* is a function $d : \{0,1\}^* \to [0, \infty)$ such that for all $w \in \{0,1\}^*$,

$$d(w) = 2^{-s}[d(w0) + d(w1)].$$

An $s$-gale can be interpreted as an strategy for betting on the successive bits of a binary string. The fairness of the gambling game depends on $s$. The notion of success corresponds to getting unbounded capital in this game.

**Definition.** Let $d : \{0,1\}^* \to [0, \infty)$ be an $s$-gale,

1. $d$ *succeeds* on a language $L \subseteq \{0,1\}^*$ if

$$\limsup_{n \to \infty} d(L[0 \ldots n-1]) = \infty.$$

2. The *success set* of $d$ is

$$S^\infty[d] = \{L \subseteq \{0,1\}^* \mid d \text{ succeeds on } L\}.$$

Effective dimension of a set is defined as the infimum $s$ for which exists an $s$-gale that succeeds on it. Depending on the computational resource that is allowed in the computation of the $s$-gale, different types of effective dimension can be defined.

**Definition.** Let $X \subseteq \mathbf{C}$ and $\Delta \in \{\text{p}, \text{p}_2, \text{pspace}, \text{plogon}\}$. The $\Delta$-*dimension of* $X$ is

$$\dim_\Delta(X) = \inf\{s \mid \text{ there exists a } \Delta\text{-computable } s\text{-gale } d \text{ s.t. } X \subseteq S^\infty[d]\}.$$

The choice of resource bounds p, $\text{p}_2$, pspace and plogon is not arbitrary, since they are suitable for quantitative study of important time and space bounded complexity classes (exponential time, exponential space, and polynomial space, respectively) and have also natural connections to information theory resource-bounded notions [15, 24].

## 2.2 Learning Models

Following [5], information is codified by the value of $n$ boolean attributes. A string in $\{0,1\}^n$ is called an *instance* and the set $\{0,1\}^n$ is called the *instance space*. A *concept* $c$ is defined as a subset of $\{0,1\}^n$ or equivalently, as a boolean function $c : \{0,1\}^n \to \{0,1\}$, where $c(x) = 1$ if $x \in c$ and $c(x) = 0$ if $x \notin c$.

Let $\mathcal{C}_n$ be a subset of concepts in the instance space $\{0,1\}^n$. A *representation* for $\mathcal{C}_n$ consists of a set of strings $L_n$ and a mapping $\sigma_n$ from $L_n$ to $\mathcal{C}_n$ that associates each string in $L_n$ with a concept in $\mathcal{C}_n$. A concept complexity measure for $\mathcal{C}_n$ is a mapping $\mathbf{size}_n$ from $\mathcal{C}_n$ to $\mathbb{N}$ (usually the minimum length of the string representing the concept in the representation $(L_n, \sigma_n)$).

For each $n \in \mathbb{N}$, let $\mathcal{C}_n$ be a set of concepts on $\{0,1\}^n$, $L_n$ and $\sigma_n$ be a representation for $\mathcal{C}_n$, and $\mathbf{size}_n$ be a concept complexity measure for $\mathcal{C}_n$. Then $\mathcal{C} = \{\mathcal{C}_n\}_n$ denotes a concept class and $\{(\mathcal{C}_n, L_n, \sigma_n, \mathbf{size}_n)\}_{n \in \mathbb{N}}$ is called the *representation class of* $\mathcal{C}$. Usually the representation and the concept complexity measure will be understood from context.

**Example 2.1** Consider the concept class $k$-CNF from Valiant's original paper [27], for some fixed $k \in \mathbb{N}$. Here the representation language $L_n$ consists of all CNF expressions on $n$ variables (say $x_1 \ldots x_n$) that have at most $k$ literals per clause, $\mathcal{C}_n$ consists of all $c \subseteq \{0,1\}^n$ such that $c$ is the set of satisfying assignments of one of these expressions, $\sigma_n$ maps a CNF expression to its set of satisfying assignments, and $\mathbf{size}_n(c)$ is the number of literals in the smallest $k$-CNF representation of $c$.

Once the notion of a concept class and its representation have been formalized, the definition of the learning models used in this paper is given next.

### 2.2.1 The Probably Approximately Correct Learning Model (PAC Learning)

The probably approximately correct (PAC) learning model formalizes the process of learning from examples. For instance, Support Vector Machines, Neural Networks and Decision Trees are based in this theoretical model.
In the PAC learning model, the learning algorithm has access to a source of positive and negative examples of an unknown target concept $c$ from a fixed and known concept class $\mathcal{C}$. The learning algorithm must approximate the target concept from the examples it has seen. Valiant formalizes this notion in [27].

More formally, let $\mathcal{C}$ and $\mathcal{H}$ be two concept classes. A *Probably Approximately Correct algorithm* for learning $\mathcal{C}$ by $\mathcal{H}$ is an algorithm that when given examples of some concept $c \in \mathcal{C}$ will produce as output (a representation) of some concept $h \in \mathcal{H}$ that is an approximation of $c$ in a sense made precise below. The class $\mathcal{C}$ is called the *target class* and $\mathcal{H}$ the *hypothesis class*. Equivalently, $c$ is called the *target concept* and $h$ the *hypothesis of the algorithm*.

**Definition.** Let $\{0,1\}^n$ be the instance space, let $D$ be a distribution on it and let $c$ be the target concept. The *error of $h$ with respect to the concept $c$ and the distribution $D$* is defined as

$$\text{error}_D(h,c) = Pr_{x \in D}[h(x) \neq c(x)].$$

That is, $\text{error}_D(h,c)$ is the probability that $h$ and $c$ do not match in a randomly chosen instance according with $D$. Intuitively, $h$ will be a good approximation of target $c$ is $\text{error}_D(h,c)$ is small.

**Definition.** Let $\mathcal{C} = \{\mathcal{C}_n\}_{n \in \mathbb{N}}$ be a concept class and let $\mathcal{H} = \{\mathcal{H}_n\}_{n \in \mathbb{N}}$ be the hypothesis class. $\mathcal{C}$ is *PAC learnable in terms of $\mathcal{H}$* if there exists a learning algorithm $A$ such that,

· for all $n \in \mathbb{N}$,

· for every target concept $c \in \mathcal{C}_n$,

· for every probability distribution $D$ on the instance space $\{0,1\}^n$,

· for all $\epsilon$ and $\delta$, where $0 < \epsilon, \delta < 1$,

if algorithm $A$ on input $(n, \epsilon, \delta)$ is given independent random examples of $\{0,1\}^n$ drawn according to $D$, together with the information of wether each example is in $c$, then with probability at least $1 - \delta$, $A$ returns a hypothesis $h \in \mathcal{H}_n$ with $\text{error}_D(h,c) \leq \epsilon$.

Moreover, the running time of $A$ is bounded by a polynomial in $n$, $1/\epsilon$, $1/\delta$ and $\textbf{size}_n(c)$.

$\mathcal{C}$ is *properly PAC learnable* if $\mathcal{C}$ is learnable in terms of $\mathcal{C}$ and $\mathcal{C}$ is *PAC learnable* if $\mathcal{C}$ is PAC learnable in terms of some class $\mathcal{H}$.

The idea beyond of this definition is that the learning algorithm must process the examples in polynomial time, i.e. be computationally efficient and must be able to produce a good approximation to the target concept with high probability using only a reasonable number of examples.

Efficiency of the learning algorithm is measured with respect to relevant parameters: size of examples ($n$), size of target concept ($\mathbf{size}_n$), $1/\epsilon$, and $1/\delta$ (see [9, 4, 5] for more details).

Notice that the definition above involves representations. It is clear that for the same concept there can be very different representations. In particular, there might be representations with very different sizes, and using a representation less succinct than other will take more time, if only to output it. Since the running time of the PAC learning algorithm is polynomial in the representation size, concepts with large representation sizes such as $\Theta(2^n)$ are trivially learnable in most cases.

**Example 2.2** [9] A number of fairly sharp results have been found for the notion of proper PAC learnability. The following summarizes some of these results. The negative results are based on the complexity of theoretic assumption that RP$\neq$ NP [17].

1. Conjunctive concepts are properly PAC learnable [27], but the class of concepts in the form disjunction of two conjunctions is not properly PAC learnable [17], and neither is the class of existential conjunctive concepts on structural instance spaces with two objects.

2. Linear threshold concepts (perceptrons) are properly PAC learnable on both Boolean and real-valued instance spaces [2], but the class of concepts in the form of the conjunction of two linear threshold concepts is not PAC learnable [1].

3. Linear thresholds of linear thresholds (i.e. multilayer perceptrons with hidden units) are properly PAC learnable, but the class of concepts in the form of the disjunction of two multilayer perceptrons is not PAC learnable. In addition, if the weights are restricted to 1 and 0 (but the threshold is arbitrary), then linear threshold concepts on Boolean instances spaces are not PAC learnable [17].

4. The classes of $k$-DNF, $k$-CNF and $k$-decision lists are properly PAC learnable for each fixed $k$ [26, 25].

Most of the difficulties in proper PAC learning are due to the computational difficulty of finding a hypothesis in the particular form specified by the target class. For example, while Boolean threshold functions with $0-1$ weights are not PAC learnable on Boolean instance spaces (unless RP=NP), they are PAC learnable by general Boolean threshold functions. Similar extended hypothesis spaces can be found for the two classes mentioned in 1, above that are not properly PAC learnable. Hence, it turns out that these classes are PAC learnable [17, 8].

### 2.2.2 The Query Learning Model

In the PAC learning model, the learning algorithm can be seen as passive, in the sense that it does not decide which examples it will see during the training phase (the labeled examples are given randomly according to some fixed distribution). However it might be interesting to allow the learning algorithm to select some particular example and ask for its labeling. This is the idea of a membership query, introduced in the seminal paper of Valiant [27] and formally defined as follows:

- *Membership query*, Mem($x$): the input is an assignment $x \in \{0,1\}^n$ and the output is the value of the target concept $c$ evaluated in $x$.

However, there are others types of queries that might be useful for the learning algorithm as those introduced in Angluin's model (query learning model) [3]. In the original definition, the learning algorithm has access a fixed set of oracles (experts) that will answer specific kinds of queries about the target concept $c$. The types of queries that are consider are the following: membership, equivalence, subset, superset, disjointness and exhaustiveness. This paper will focus in learning algorithms that have access to membership queries or equivalence queries.
An equivalence query is formally defined as:

- *Equivalence query*, Equ($h$): the input is a representation of a concept $h \in \mathcal{H}_n$ and the output is either YES, if $h$ is equivalent to the target concept $c$, or NO indicating that they are not equivalent. In the latter case, a *counterexample* $x$ satisfying $c(x) \neq h(x)$, is also returned.

Thus, learning with membership or equivalence queries is formally defined as follows.

**Definition.** Let $\mathcal{C} = \{\mathcal{C}_n\}_{n \in \mathbb{N}}$ be a concept class and let $\mathcal{H} = \{\mathcal{H}_n\}_{n \in \mathbb{N}}$ be the hypothesis class. $\mathcal{C}$ is *learnable in terms of $\mathcal{H}$ from membership (equivalence) queries* if there exists an algorithm $A$ such that, for every $n$ and every concept $c \in \mathcal{C}_n$, $A$ asks membership (equivalence) queries about $c$ and eventually halts and output a hypothesis $h \in \mathcal{H}_n$ that is equivalent to the target, i.e. for all $x$, $c(x) = h(x)$.

$\mathcal{C}$ is *efficiently learnable from membership or equivalence queries* if the running time and the total number of queries made by $A$ are bounded by a polynomial in $n$ and $\textbf{size}_n(c)$.

Notice again that the choice of representation is very relevant for query learnability, in particular representation size. A concept with a representation size $\Theta(2^n)$ is trivially learnable in most settings.

### 2.2.3 The On-line Mistake-bound Learning Model

This learning model is the on-line mistake-bound model of Littlestone [19], that considers learning from examples in a situation in which the goal of the learner is simply to make few mistakes. In this model, there is no separate set of training examples. The learner attempts to predict the appropriate response for each example (i.e. predict if it is a negative or positive example), starting with the first example received. After making this prediction, the learner is told whether the prediction was correct, and then uses this information to improve its hypothesis. The learner continues to learn as long as it receives examples; that is, it continues to examine the information it receives in an effort to improve its hypothesis. The evaluation of the algorithm's learning behavior is made by counting the worst-case number of mistakes that it will make while learning a concept from a specified concept class. This corresponds to learning in an adversary setting for the order of the examples received.

**Definition.** Let $\mathcal{C} = \{\mathcal{C}_n\}_{n \in \mathbb{N}}$ be a concept class,

1. An *on-line learning algorithm $A$* is an algorithm which gets inputs of the form $(n, y, h)$ and outputs 0 or 1, where $|y| = n$ is the example to be predicted and $h = ((x_1, b_1) \ldots (x_r, b_r))$ is the history of previously received examples together with the corresponding correct answers.

2. For an integer $n$, and a concept $c \in \mathcal{C}_n$, the *number of mistakes made by $A$ on $c$* is defined as follows,

$$\text{Mist}(n, c, A) = \max_h \#\{x \mid |x| = n, A(n, x, h) \neq c(x)\}.$$

3. The *worst-case number of mistakes made by $A$ of $\mathcal{C}_n$* is defined as

$$\text{Mist}(n, \mathcal{C}_n, A) = \max_{c \in \mathcal{C}_n} \text{Mist}(n, c, A).$$

4. The concept class $\mathcal{C}$ is *on-line learnable with $f(n)$ mistakes* if there exists an on-line algorithm $A$, which runs in time polynomial in its input length, so that for infinitely many $n$, $\text{Mist}(n, \mathcal{C}_n, A) \leq f(n)$.

**Proposition 2.3** *[19] If $\mathcal{C}$ is learnable with $f(n)$ equivalence queries, then $\mathcal{C}$ is on-line learnable with $f(n)$ mistakes.*

Notice that the time bounds of the learning algorithms are relevant in proposition 2.3.

## 2.3 Dimension of concept classes

In this paper, a concept class $\mathcal{C}$ will have two representations:

- The representation $(L_n, \sigma_n)$ used by the learning algorithm.

- The representation given by the characteristic sequence for each length, that is, for $L \in \mathcal{C}$, $n \in \mathbb{N}$, $L^{=n}$ is a string of length $2^n$. This representation will be used to define the dimension of a concept class.

Notice that with this convention, the dimension of a concept class does not depend on the representation used for each individual concept. That is, we will assume the existence of some representation $\sigma_n$ with $\textbf{size}_n$ bounded in terms of $n$ and the existence of some learning algorithm, but the results will not be focused on the particular representation.

Also, we will assume that there exist an hypothesis class $\mathcal{H}$ from which $\mathcal{C}$ is learnable, but the results in this paper will not depend on the particular $\mathcal{H}$. Thus, we will abbreviate *learnable in terms of $\mathcal{H}$* by *learnable*.

# 3 Dimension and on-line mistake-bound learning

This section focuses on the on-line mistake-bound learning model. We prove that the mistake bound of on-line learning provides an upper bound on the p-dimensions of each class. Moreover, this upper bound can be tight for certain classes.

The relationship of dimension with logarithmic loss unpredictability was explored in [12] and is intuitively close to on-line learning when restricting to examples given in lexicographical order.

Based on the results in this section (although with chronologically earlier publications) Hitchcock [14] has further explored the case of dimension zero and small mistake bounds for on-line learning, including reductions to these classes.

The main theorem in this section provides an upper bound of the p-dimension for concept classes that are on-line learnable with $\alpha 2^n$ mistakes.

**Theorem 3.1** *Let $\alpha \leq 1/2$ be a p-computable number. Let $\mathcal{C}$ be a concept class that is learnable with $\alpha 2^n$ mistakes, then*

$$\dim_p(\mathcal{C}) \leq \mathcal{H}(\alpha),$$

*where $\mathcal{H}$ is Shannon binary entropy, $\mathcal{H}(\alpha) = \alpha \log \frac{1}{\alpha} + (1-\alpha) \log \frac{1}{1-\alpha}$.*

**Proof.** Let $\alpha < 1/2$ (the case $\alpha = 1/2$ is trivial). We will show that for any $s > \mathcal{H}(\alpha)$, there exists an $s$-gale that succeeds on $\mathcal{C}$. Let $\epsilon = \frac{s - \mathcal{H}(\alpha)}{2}$.

We will use the following function $h_\alpha(x) = \alpha \log \frac{1}{x} + (1-\alpha) \log \frac{1}{1-x}$. This is a continuous function in the range $(0, 1)$ which takes the minimum value $\mathcal{H}(\alpha)$ at $x = \alpha$. Let $\delta$ be such that $h_\alpha(\alpha + \delta) \leq \mathcal{H}(\alpha) + \epsilon$, and $\alpha + \delta < 1/2$.

Let $A$ be an algorithm that learns $\mathcal{C}$ with $\alpha 2^n$ mistakes. For each $z \in \{0,1\}^*$ of length between 0 and $2^n$ we denote by $h(z)$ the history corresponding to having received examples $s_0^n \ldots s_{|z|-1}^n$ with corresponding correct answers $z[0] \ldots z[|z|-1]$, that is, examples in lexicographical order and answers recorded in the bits of $z$. We define an $s$-gale $d : \{0,1\}^* \to [0, \infty)$ recursively as follows.

$$d(\lambda) = 1.$$

Let $n \in \mathbb{N}$, for any $w$ with $2^n - 1 \leq |w| < 2^{n+1} - 1$, we define

$$d(wb) = \begin{cases} (\alpha + \delta) 2^s d(w) & \text{if } A(n, h(w[2^n - 1 \ldots |w| - 1]), s_{|w|}) = \bar{b}, \\ (1 - (\alpha + \delta)) 2^s d(w) & \text{if } A(n, h(w[2^n - 1 \ldots |w| - 1]), s_{|w|}) = b. \end{cases}$$

Notice that $d$ can be computed in polynomial time since $A$ works in time polynomial in the input length (where the input of $A$ includes history).

Let $L \in \mathcal{C}$ be a concept. Then,

$$\begin{aligned} d(L[0 \ldots 2^{n+1} - 2]) &= d(L^{=0} \ldots L^{=n}) \\ &\geq \left[ (\alpha + \delta)^{\text{Mist}(n, L^{=n}, A)} (1 - (\alpha + \delta))^{2^n - \text{Mist}(n, L^{=n}, A)} \right] 2^{2^n s} d(L^{=0} \ldots L^{=n-1}) \\ &\geq \left[ (\alpha + \delta)^{\alpha 2^n} (1 - (\alpha + \delta))^{(1-\alpha) 2^n} \right] 2^{2^n s} d(L^{=0} \ldots L^{=n-1}) \\ &= 2^{-h_\alpha(\alpha + \delta) 2^n} 2^{2^n s} d(L^{=0} \ldots L^{=n-1}) \\ &= 2^{2^n (s - h_\alpha(\alpha + \delta))} d(L^{=0} \ldots L^{=n-1}) \\ &\geq 2^{2^n (s - \mathcal{H}(\alpha) - \epsilon)} d(L^{=0} \ldots L^{=n-1}) \\ &\geq 2^{\sum_{i=0}^n 2^i (s - \mathcal{H}(\alpha) - \epsilon)} = 2^{(2^{n+1} - 1)\epsilon} \end{aligned}$$

that tends to infinity with $n$. Therefore $\mathcal{C} \subseteq S^\infty[d]$ and $\dim_p(\mathcal{C}) \leq s$. □

As a corollary classes of concepts on-line learnable with $o(2^n)$ mistakes have p-dimension 0. This corollary was later generalized by Hitchcock [14].

**Corollary 3.2** *Let $\mathcal{C}$ be a class of concepts on-line learnable with at most $o(2^n)$ mistakes. Then,*
$$\dim_p(\mathcal{C}) = 0.$$

The following corollary is an immediate consequence of Proposition 2.3 and improves a result by Lindner, Schuler and Watanabe [18].

**Corollary 3.3** *If Boolean circuits are learnable in polynomial time (even linear exponential time) with $o(2^n)$ equivalences queries then the concept class of Boolean circuits has p-dimension 0.*

Next we prove that Theorem 3.1 is optimal by presenting a concept class that is on-line learnable with $\alpha 2^n$ mistakes and has p-dimension $\mathcal{H}(\alpha)$.

**Theorem 3.4** *Let $\alpha \leq 1/2$ be a p-computable number. There exists a concept class $\mathcal{C}_\alpha$ that is on-line learnable with $\alpha 2^n$ mistakes such that $\dim_p(\mathcal{C}_\alpha) = \mathcal{H}(\alpha)$.*

**Proof.** For each $\alpha$ consider the concept class
$$\mathcal{C}_\alpha = \{L \in \mathbf{C} \mid \forall n, \#L^{=n} \leq \alpha 2^n\}$$
and in line with the proof of Lemma 5.1. [22], we can show that $\dim_p(\mathcal{C}_\alpha) = \mathcal{H}(\alpha)$. Consider the algorithm $A$ which predicts 0 all the time. The number of mistakes made by this algorithm on any concept in $\mathcal{C}_\alpha$ is at most $\alpha 2^n$.
□

Our last result shows that the values of the dimension of on-line learnable classes with $\alpha 2^n$ mistakes are dense in the interval $[0, \mathcal{H}(\alpha)]$.

**Theorem 3.5** *Let $\alpha \leq 1/2$ be a p-computable number and let $\beta \in [0, \mathcal{H}(\alpha))$ be p-computable. Then, there exists a concept class $\mathcal{C}_\beta$ that is on-line learnable with $\alpha 2^n$ mistakes such that*

$$\dim_p(\mathcal{C}_\beta) = \beta.$$

**Proof.** Let $\beta > 0$ and let $\mathcal{C}_\beta = \{L \in \mathbf{C} \,|\, \forall n, \#L^{=n} \leq \gamma 2^n\}$, where $\gamma$ is the smallest value such that $\mathcal{H}(\gamma) = \beta$. Notice that, $\mathcal{H}(x)$ is a symmetric and an strictly increasing continuous function for $x \leq 1/2$, so $\gamma \leq \alpha$. By Theorem 3.4, $\dim_\mathrm{p}(\mathcal{C}_\beta) = \beta$ and $\mathcal{C}_\beta$ is on-line learnable with $\gamma 2^n$ mistakes, so it is on-line learnable with $\alpha 2^n$ mistakes.

The case $\beta = 0$ holds trivially with the same definition of $\mathcal{C}_\beta$. □

As the last remark, we can generalize all results in this section by using a weaker on-line learning model that is only required to learn when examples are given on lexicographical order.

## 4 Dimension and PAC Learning

This section is focused on the PAC learning model. This model was related to resource-bounded measure [21] by Watanabe et al [18]. In this section we show that it is also related with polynomial space dimension, partially generalizing [18].

Our main result here proves that polynomial-space dimension of concept classes that are learnable by a PAC algorithm is zero. This result can be used to demonstrate that a large class $\mathcal{C}$ (in the dimension sense) is not efficiently PAC learnable using any hypothesis class $\mathcal{H}$ (that is, in notation of [16], $\mathcal{C}$ is inherently unpredictable).

Finally we show a stronger result for plogon and $p_2$ as dimension resource bounds, but in this case, some extra hypotheses are required.

Since the PAC learner resource bounds (time and space) depend on the representation size, only classes with subexponential representations ($\mathbf{size}_n(c) \in o(2^n) \forall c$) have a real interest.

**Theorem 4.1** *Let $\mathcal{C}$ be a PAC learnable concept class with subexponential representations. Then*
$$\dim_{\mathrm{pspace}}(\mathcal{C}) = 0.$$

*Moreover, if there exists a PAC algorithm that runs in space $O(2^n)$ with a number of examples $\xi(n)$ verifying $\sum_{i=0}^n \xi(i) \in o(2^n)$, then*

$$\dim_{\mathrm{pspace}}(\mathcal{C}) = 0.$$

**Proof.** The first statement is a particular case of the second one, so it is enough to prove the latter.

Let $A$ be the PAC algorithm that witnesses that $\mathcal{C}$ is PAC learnable. Let $D$ be the uniform distribution. Let $s > \mathcal{H}(\epsilon)$. Let $c \in \mathcal{C}_n$. Then, the algorithm $A$ on input $(n, \epsilon, \delta)$ outputs (with probability $1 - \delta$) a hypothesis $h$ such that $h$ $\epsilon$-approximates $c$. Then, with probability $1 - \delta$,

$$\frac{\#\{x \in \{0,1\}^n | h(x) \neq c(x)\}}{2^n} = \mathrm{err}_D(c, h) \leq \epsilon.$$

Let $Q_n$ be the class of possible sets of examples that $A(n, \epsilon, \delta)$ can use, i.e.

$$Q_n = \{Q \subseteq \{0,1\}^n \mid \#Q \leq \xi(n)\}.$$

Now, let $Q \in Q_n$ and $w$ of length $2^n$ (that is the lexicographical representation of a concept $c \in \mathcal{C}_n$). Then, we say that $w$ *is good for $A$ with respect to $Q$* if

$$A^{c,Q}(n, \epsilon, \delta) = h \text{ with } \mathrm{err}_D(c, h) \leq \epsilon,$$

where the notation $A^{c,Q}$ references the output of $A$ when $A$ is given as examples the elements of $Q$ together with information of whether they are or not in $c$.

Intuitively, $w$ is good for $A$ with respect to $Q$ if we can learn approximately the concept $c$ represented by $w$ using the examples that $Q$ provides to $A$.

Let $B_{n,Q}$ be the set of sequences of length $2^n$ that are good for $A$ with respect to $Q$ and let $d_{n,Q} : \{0,1\}^{\leq 2^n} \to [0, \infty)$ be the function defined as follows,

$$d_{n,Q}(v) = \frac{\#\{w \text{ good for } A \text{ with respect to } Q \mid v \sqsubseteq w\}}{\#B_{n,Q}}.$$

Notice that, by reusing space, $d_{n,Q}$ is computable in space $O(2^n)$. Next, the following function is defined by considering all $Q \in Q_n$,

$$d_n(v) = \frac{\sum_{Q \in Q_n} d_{n,Q}(v)}{\#Q_n}.$$

Notice that $d_n$ is also computable in space $O(2^n)$. Moreover,

i) $d_n(\lambda) = 1$.

ii) $d_n$ verifies $d_n(w0) + d_n(w1) = d_n(w)$ for all $|w| < 2^n$.

iii) If $w$ is a sequence of length $2^n$ then

$$d_n(w) = \sum_{Q \in Q_n(w)} \frac{1}{\#Q_n \cdot \#B_{n,Q}}, \tag{1}$$

where $Q_n(w)$ is defined by

$$Q_n(w) = \{Q \in Q_n \mid w \text{ is good for } A \text{ with respect to } Q\}.$$

Now, we define the function $d : \{0,1\}^* \to [0, \infty)$ as

$$d(w) = 2^{s|w|} \prod_{i=0}^{n} d_i(w^i),$$

where $w = w^0 \ldots w^n$ with $|w^i| = 2^i$ for all $0 \leq i < n$ and $|w^n| \leq 2^n$.

It's easy to see that $d$ is an $s$-gale. Also, since each $d^i$ is computable in space $O(2^i)$ (with $i \leq n$) and $i \leq \log |w|$, $d \in$ pspace.

Finally, we will be need the following lemma that gives an upper bound on the number of sequences that are good for $A$ with respect to any $Q$.

**Lemma 4.2** *For all $n \in \mathbb{N}$ and $Q \in Q_n$ we have that*

$$\#B_{n,Q} \leq 2^{\mathcal{H}(\epsilon)2^n} 2^{\xi(n)},$$

*where $\epsilon$ is the error parameter in the PAC algorithm $A$.*

**Proof.** Let us see how many different hypotheses the algorithm $A$ can return when using the examples provided by a fixed set of examples $Q$. Notice that each $Q \in Q_n$ verifies that $\#Q \leq \xi(n)$, thus $A$ can generate at most $2^{\xi(n)}$ hypotheses.

Fix one of those hypotheses, say $h$, and let $\tilde{h} \in \{0,1\}^{2^n}$ be its characteristic sequence. Then, we estimate the number of sequences that are an $\epsilon$-approximation of $\tilde{h}$ as follows. If

$$Approx(\epsilon, \tilde{h}) = \{w \in \{0,1\}^{2^n} \mid \#\{i \in \{0 \ldots 2^n - 1\} \mid \tilde{h}[i] \neq w[i]\} \leq \epsilon 2^n\},$$

then by Chernoff bound [6]

$$\#Approx(\epsilon, \tilde{h}) = \sum_{k=0}^{\epsilon 2^n} \binom{2^n}{k} \leq 2^{\mathcal{H}(\epsilon)2^n}.$$

So, for each hypothesis $h$ there are at most $2^{\mathcal{H}(\epsilon)2^n}$ sequences that $\epsilon$-approximate $\tilde{h}$ and thus, for each $n \in \mathbb{N}$,

$$\#B_{n,Q} \leq 2^{\mathcal{H}(\epsilon)2^n} 2^{\xi(n)}.$$

$\square$

Now, let us see that $d$ succeeds on $\mathcal{C}$. Let $L \in \mathcal{C}$ and let $c \in \mathcal{C}_n$ be the concept represented by $L^{=n}$. On the one hand, since $A(n, \epsilon, \delta)$ returns with probability $1 - \delta$ an $\epsilon$-approximation of $c$, we have that

$$\frac{\#Q_n(L^{=n})|}{\#Q_n} \geq 1 - \delta. \tag{2}$$

On the other hand, by (1)

$$d_n(L^{=n}) = \sum_{Q \in Q_n(L^{=n})} \frac{1}{\#Q_n \cdot \#B_{n,Q}}.$$

Using Lemma 4.2 in the last equation we have that

$$d_n(L^{=n}) \geq \sum_{Q \in Q_n(L^{=n})} \frac{1}{\#Q_n 2^{\mathcal{H}(\epsilon)2^n} 2^{\xi(n)}}$$
$$\geq \frac{1-\delta}{2^{\mathcal{H}(\epsilon)2^n} 2^{\xi(n)}},$$

where the last inequality is obtained using (2).

Therefore, for all $n \in \mathbb{N}$,

$$d(L[0\ldots 2^{n+1}-2]) \geq 2^{s(2^{n+1}-1)} \prod_{i=0}^{n} \frac{1-\delta}{2^{\mathcal{H}(\epsilon)2^i} 2^{\xi(i)}}$$
$$= 2^{s(2^{n+1}-1)} \frac{(1-\delta)^{n+1}}{2^{\sum_{i=0}^{n} \mathcal{H}(\epsilon)2^i + \xi(i)}}$$
$$= 2^{(s-\mathcal{H}(\epsilon))(2^{n+1}-1)} \frac{(1-\delta)^{n+1}}{2^{\sum_{i=0}^{n} \xi(i)}},$$

that tends to infinity since $s > \mathcal{H}(\epsilon)$. Finally, $\epsilon > 0$ is arbitrary and $\mathcal{H}(\epsilon)$ tends to 0 when $\epsilon \to 0$ so, for all $s > 0$, we can define an $s$-gale in pspace that succeeds on $\mathcal{C}$. □

Notice that the above theorem is true for any hypothesis class that we may consider. So, all the positive results in Example 2.2 (both PAC learnable and properly PAC learnable) can be used to obtain results on pspace-dimension.

**Corollary 4.3** *The following classes have polynomial-space dimension zero:*

1. *The class of conjunctive concepts.*

2. *Linear threshold concepts (perceptrons). In fact, it is proven in [7] that this class has also polynomial-time dimension zero.*

3. *The class of concepts in the form linear thresholds of linear thresholds (i.e. multilayer perceptrons with hidden units).*

4. *The classes of k-DNF, k-CNF and k-decision lists (for each fixed k).*

Theorem 4.1 can also be used in a negative way, obtaining the following strong nonlearnability result for any hypothesis class of a concept class.

**Corollary 4.4** *Let $\mathcal{C}$ be a concept class such that $\dim_{\text{pspace}}(C) \neq 0$. Then, $\mathcal{C}$ is inherently unpredictable, i.e. there is no class of hypothesis for which $\mathcal{C}$ is PAC learnable.*

We can generalize Theorem 4.1 to PAC algorithms that use a larger number of examples.

**Theorem 4.5** *Let $\mathcal{C}$ be a concept class that can be learned by a PAC algorithm in space $O(2^n)$ with at most $\alpha 2^n$ examples ($\alpha \leq 1$ pspace-computable), then*
$$\dim_{\text{pspace}}(\mathcal{C}) \leq \alpha.$$

**Proof.** The proof is analogous to the last theorem, just using $\sum_{i=0}^{n} \xi(i) \leq \alpha(2^{n+1} - 1)$. Thus
$$d(L[0 \ldots 2^{n+1} - 2]) \geq 2^{(s-\mathcal{H}(\epsilon)-\alpha)(2^{n+1}-1)}(1-\delta)^{n+1},$$

that tends to infinity when $s > \mathcal{H}(\epsilon) + \alpha$. Finally, $\epsilon > 0$ is arbitrary and $\mathcal{H}(\epsilon)$ tends to 0 when $\epsilon \to 0$. So, for all $s > \alpha$ we can define an $s$-gale in pspace such that succeeds on $\mathcal{C}$. □

Finally we look at more efficient dimension versions and obtain the following.

**Theorem 4.6** *Let $\mathcal{C}$ be a concept class that can be learned by a PAC algorithm with working space and number of examples bounded by $p(n)$, for $p$ a fixed polynomial. Then*
$$\dim_{\text{plogon}}(\mathcal{C}) = 0.$$

**Theorem 4.7** *Let $\mathcal{C}$ be a concept class that can be learned by a PAC algorithm within time $2^n$ and with number of examples bounded by $p(n)$, for $p$ a fixed polynomial. Then*
$$\dim_{\text{p2}}(\mathcal{C}) = 0.$$

Notice that the polynomial bounds in the theorems above do not imply a trivial representation size for either the concept class or the hypothesis, since the PAC algorithm output space is not likewise bounded.

# 5  Dimension and membership-query learning

This section is focused on the membership-query learning model. The main result proves that polynomial-space dimension of concept classes that are learnable by a membership-query algorithm that runs in space $O(2^n)$ and makes at most $o(2^n)$ queries is zero. This implies that large classes in the dimension sense require long representations in order to be membership learnable. Finally we show a stronger result under more restricted learnability conditions.

**Theorem 5.1** *Let $\mathcal{C}$ be a class of concepts learnable with a membership-query algorithm that runs in space $O(2^n)$ and makes at most $o(2^n)$ queries, then*
$$\dim_{\text{pspace}}(\mathcal{C}) = 0.$$

**Proof.**

Let $A$ be the query learning algorithm that witnesses that $\mathcal{C}$ is learnable with $o(2^n)$ membership queries. Let $q(n)$ be the maximum number of queries of $A$ on input $n$.

Let $w$ be a string of length $2^n$. We say that $w$ is good for $A$ if the algorithm $A$ on input $n$ outputs a hypothesis $h$ equivalent to $w$ and the total number of queries made by $A$ is bounded by $q(n)$. Let $B_n$ be the set of all sequences of length $2^n$ that are good for $A$.

Notice that, if only membership queries are allowed, then the number of different outputs of $A(n)$ is bounded by $2^{q(n)}$, so $\#B_n \leq 2^{q(n)}$.

Let $d_n : \{0,1\}^{\leq 2^n} \to [0, \infty)$ be the function defined as follows,

$$d_n(v) = \frac{\#\{w \text{ good for } A \mid v \sqsubseteq w\}}{\#B_n}.$$

Notice that, by reusing space, $d_n$ can be computed in space $O(2^n)$. Also, if $w$ has length $2^n$ and is good for $A$, $d_n(w) = 1/\#B_n$.

Now, the $s$-gale $d : \{0,1\}^* \to [0, \infty)$ is defined by

$$d(w) = 2^{s|w|} \prod_{i=0}^{n} d_i(w^i),$$

where $w = w^0 \ldots w^n$ with $|w^i| = 2^i$ for all $0 \leq i < n$ and $|w^n| \leq 2^n$.

It's easy to see that $d$ is an $s$-gale. Also, since each $d_i$ is computable in space $O(2^i)$ (with $i \leq n$) and $i \leq \log |w|$, $d \in \text{pspace}$.

Finally, let us prove that $d$ succeeds on $\mathcal{C}$. Let $L \in \mathcal{C}$ and $n \in \mathbb{N}$, then $L^{=n}$ is good for $A$ and

$$d_n(L^{=n}) \geq \frac{1}{\#B_n} \geq \frac{1}{2^{q(n)}}.$$

Thus, for all $n \in \mathbb{N}$,

$$d(L[0\ldots 2^{n+1}-2]) \geq 2^{s(2^{n+1}-1)} \prod_{i=0}^{n} \frac{1}{2^{q(i)}}$$

that tends to infinity when $s > 0$. $\square$

If the number of queries in Theorem 5.1 is allowed to be up to $\alpha 2^n$, then $\alpha$ is an upper bound for polynomial-space dimension of $\mathcal{C}$.

**Theorem 5.2** *Let $\mathcal{C}$ be a class of concepts learnable with a membership-query algorithm that runs in space $O(2^n)$ and makes at most $\alpha 2^n$ queries ($\alpha \leq 1$, $\alpha$ pspace computable), then*

$$\dim_{\text{pspace}}(\mathcal{C}) \leq \alpha.$$

**Proof.** The proof is analogous to the above theorem, just consider $q(n) = \alpha 2^n$. In this case,
$$\#B_n \leq 2^{\alpha 2^n},$$
and then,

$$\begin{aligned} d(L[0\ldots 2^{n+1}-2]) &\geq 2^{s(2^{n+1}-1)} \prod_{i=0}^{n} \frac{1}{2^{\alpha 2^i}} \\ &= 2^{(s-\alpha)(2^{n+1}-1)}. \end{aligned}$$

that tends to infinite when $s > \alpha$. Therefore, $\dim_{\text{pspace}}(\mathcal{C}) \leq \alpha$. $\square$

The following theorem shows that Theorem 5.2 is optimal.

**Theorem 5.3** *Let $\alpha \in \mathbb{Q} \cap (0,1)$. There exists a concept class $\mathcal{C}_\alpha$ that is query-learnable with $\alpha 2^n$ membership queries such that*

$$\dim_{\text{pspace}}(\mathcal{C}_\alpha) = \alpha.$$

**Proof.** We will use a construction from Theorem 4.3. in [23]. Let $L \in \mathbf{C}$ and let $L = L_1 L_2 L_3 \ldots$ be a partition of $L$ with $|L_i| = \alpha 2^i$. We define the sequence $\tilde{L} \in \mathbf{C}$ as the concatenation of $\tilde{L}_i = L_i 0^{2^i - |L_i|}$ with $|\tilde{L}_i| = 2^i$.

Let $\mathcal{C}_\alpha = \{\tilde{L} \mid L \in \mathbf{C}\}$, then it is clear that this class is learnable with a query-learning algorithm that makes $\alpha 2^n$ queries. Notice that it is only necessary to query about the bits that are provided from the original sequences $L_i$, because the other bits are all zero, and these are exactly $\alpha 2^n$ for each $n$.

Let us see that $\dim_{\text{pspace}}(\mathcal{C}) = \alpha$. First, we will see that $\dim_{\text{pspace}}(\mathcal{C}) \leq \alpha$. Let $s \in [0,1]$ and define $d : \{0,1\}^* \to [0,\infty)$ as follows:

i) $d(\lambda) = 1$.

ii) Let $w = w_0 \ldots w_m$ with $|w_i| = 2^i$ for all $i \leq m$ and $|w_m| \leq 2^m$, then

$$d(wb) = \begin{cases} 2^{s-1} d(w) & \text{if } |w_m| < \alpha 2^m \\ 2^s d(w) & \text{if } |w_m| \geq \alpha 2^m \text{ and } b = 0. \\ 0 & \text{if } |w_m| \geq \alpha 2^m \text{ and } b = 1. \end{cases}$$

It is clear that this function is a pspace-computable $s$-gale. Let $\tilde{L} \in \mathcal{C}$, then

$$\begin{aligned} d(\tilde{L}[0 \ldots 2^n - 2]) &= d(\tilde{L}^{=0} \ldots \tilde{L}^{=n}) = \\ &= \prod_{i=0}^n 2^{(s-1)\alpha 2^i} 2^{s(1-\alpha)2^i} \\ &= 2^{\sum_{i=0}^n 2^i (s-\alpha)} = 2^{(2^{n+1}-1)(s-\alpha)} \end{aligned}$$

that tends to infinity when $s > \alpha$, so $\dim_{\text{pspace}}(\mathcal{C}) \leq \alpha$.

Let us see that $\dim_{\text{pspace}}(\mathcal{C}) \geq \alpha$ using a gale-diagonalization technique. Let $s < \alpha$ and let $d$ be a pspace-computable $s$-gale. We will recursively build a sequence $\tilde{L} \in \mathcal{C}$ such that $d$ does not succeed on $\tilde{L}$. Suppose that $\tilde{L}[0 \ldots n-1]$ has been built and let $\tilde{L}[0 \ldots n-1] = L_0 \ldots L_m$ where $|L_i| = 2^i$ and $|L_m| \leq 2^m$, we define

$$\tilde{L}[n] = \begin{cases} b & \text{if } |\tilde{L}_m| < \alpha 2^m \text{ and } d(\tilde{L}[0 \ldots n-1]b) \leq d(\tilde{L}[0 \ldots n-1]\bar{b}) \\ 0 & \text{if } |\tilde{L}_m| \geq \alpha 2^m. \end{cases}$$

It is clear that $\tilde{L} \in \mathcal{C}$, let us see now that $d$ does not succeed on $\tilde{L}$. In the best case, $d$ wins $2^s$ of the current capital in the $(1-\alpha)2^i$ lasts bits of any $\tilde{L}_i$, and loses at least $2^{s-1}$ of the capital in the other bits (this will be the case when $d(\tilde{L}[0 \ldots n-1]b) = d(\tilde{L}[0 \ldots n-1]\bar{b})$). So,

$$\begin{aligned} d[0 \ldots 2^{m-1} - 2] &= \sum_{i=0}^m 2^{s(1-\alpha)2^i} 2^{(s-1)\alpha 2^i} \\ &= \sum_{i=0}^m 2^{2^i(s-\alpha)} \end{aligned}$$

that does not tend to infinity when $s < \alpha$. □

As a corollary of Theorem 5.1, we know that large classes (in the dimension sense) require large representations in order to be membership learnable.

**Corollary 5.4** *Let $\mathcal{C}$ be a concept class such that*

$$\dim_{\text{pspace}}(\mathcal{C}) \neq 0.$$

*Then $\mathcal{C}$ does not have a representation of size $o(2^n)$ for which it is efficiently learnable by membership queries.*

**Proof.** By definition, $\mathcal{C}$ is efficiently learnable if the running time and the total number of queries made by $A$ are bounded by a polynomial in $n$ and in $\textbf{size}_n(c)$. Thus, if $\textbf{size}_n \in o(2^n)$, the running time (and then the working space) is $o(2^n)$ and $\dim_{\text{pspace}}(\mathcal{C}) = 0$, which is a contradiction. □

Finally, the following theorem proves that Theorem 5.1 is also true for plogon-dimension when polynomial bounds are required.

**Theorem 5.5** *Let $\mathcal{C}$ be a class of concepts learnable with a membership-query algorithm that runs in space polynomial in $n$ and makes at most a polynomial number of queries. Then,*

$$\dim_{\text{plogon}}(\mathcal{C}) = 0.$$

Both output space and time of the learning algorithm are not restricted in the theorem above. Therefore nontrival representation sizes can be learned under the conditions of Theorem 5.5.

The above theorem is also true for $p_2$-dimension.

## 6 Future work

Results connecting PAC-learning with p-dimension would shed new light on the learnability of languages in exponential time (EXP).